\documentclass[prd,aps,preprint,tightenlines]{revtex4}

\begin{document}

\newcommand{\TeV}{\,{\rm TeV}}
\newcommand{\GeV}{\,{\rm GeV}}
\newcommand{\MeV}{\,{\rm MeV}}
\newcommand{\keV}{\,{\rm keV}}
\newcommand{\eV}{\,{\rm eV}}
\def\ap{\approx}
\newcommand{\bea}{\begin{eqnarray}}
\newcommand{\eea}{\end{eqnarray}}
\def\beq{\begin{equation}}
\def\eeq{\end{equation}}
\def\haf{\frac{1}{2}}
\def\lpp{\lambda''}
\def\ccg{\cal G}
\def\slash#1{#1\!\!\!\!\!\!/}
\def\u{{\cal U}}
\def\inf{{\cal 1}}

\setcounter{page}{1}
\preprint{KAIST-TH 01/12, SNUTP 01-018, hep-ph/0107083}

\title{Bi-maximal neutrino mixing and small $U_{e3}$
from Abelian flavor symmetry}

\author{Kiwoon Choi$^a$, Eung Jin Chun$^b$,
Kyuwan Hwang$^a$ and Wan Young Song$^a$}

\affiliation{$^a$Department of Physics, Korea Advanced Institute of Science and 
Technology \\
        Taejon 305-701, Korea \\
$^b$School of Physics,  Seoul National University, Seoul 151-147, Korea 
}


\begin{abstract}
Atmospheric neutrino data strongly suggests a near-maximal
$\nu_\mu$--$\nu_{\tau}$ mixing and also
solar neutrino data can be nicely explained by 
another near-maximal $\nu_e$--$\nu_{\mu}$ or $\nu_e$--$\nu_{\tau}$ mixing.
We examine the possibility that this bi-maximal mixing of
atmospheric and solar neutrinos arises naturally, while
keeping $U_{e3}$ and $\Delta m^2_{\rm sol}/\Delta m^2_{\rm atm}$ small
enough, as a consequence of Abelian flavor symmetry.
Two simple scenarios of Abelian flavor symmetry within supersymmetric 
framework are considered to obtain the desired form of 
the neutrino mass matrix and the charged lepton mass matrix 
parameterized by the Cabibbo angle $\lambda\approx 0.2$.
Future experiments at a neutrino factory measuring the size 
of $U_{e3}$ and the sign of $\Delta m^2_{32}$ could discriminate 
those scenarios as they predict  distinctive values of $U_{e3}$ 
in connection with $\Delta m^2_{\rm sol}/\Delta m^2_{\rm atm}$ 
and also with the order of the neutrino mass eigenvalues.
\end{abstract}


\maketitle


\section{Introduction}

Atmospheric and solar neutrino experiments have suggested for a long
time that neutrinos oscillate into different flavors.
In particular, the Super-Kamiokande data strongly indicate
that the observed deficit of atmospheric muon neutrinos is due
to the near-maximal $\nu_\mu\rightarrow \nu_\tau$ oscillation
\cite{atm}.  Solar neutrino data from the recent SNO experiment 
combined with those of Homestake, SAGE, GALLEX and Super-Kamiokande 
\cite{solex} provide also strong observational basis for
$\nu_e\rightarrow \nu_\mu$ or $\nu_\tau$ oscillation \cite{p-sno}.
Thus, the ``standard'' framework to accomodate the atmospheric and 
solar neutrino anomalies is to introduce small but nonzero masses 
of the three known neutrino species.
 
Low energy effective Lagrangian relevant to
the neutrino masses and mixing can be written as
\beq
\Delta{\cal L}\, = \,\overline{e_L} M^{e} e_R +
  g W^{-\mu}\overline{e_L}\gamma_\mu \nu_L 
  + {1\over 2}\overline{(\nu_L)^c} M^\nu \nu_L + {\rm h.c.},
\eeq
where the charged lepton mass matrix $M^e$ and 
the neutrino mass matrix $M^\nu$ are not diagonal in general 
in the weak interaction eigenbasis.
Diagonalizing $M^e$ and $M^\nu$ as
\bea
&& (U^e)^{\dagger} M^e V^e = D^e ={\rm diag} \, (m_e,
m_\mu, m_\tau),
\nonumber \\
&&
(U^\nu)^T M^{\nu} U^\nu = D^\nu ={\rm diag}\, (m_1,m_2,m_3),
\eea
one finds the effective Lagrangian written in terms of
the mass eigenstate fermion fields:
\beq
\Delta{\cal L} \,= \, \overline{e_L} D^{e} e_R +
  g W^{-\mu}\overline{e_L}\gamma_\mu U^{\rm MNS} \nu_L 
  + {1\over 2}\overline{(\nu_L)^c} D^\nu \nu_L
  + {\rm h.c.}\,,
\eeq
where the MNS lepton mixing matrix\cite{MNS} is given
by \beq
\label{MNS}
U^{\rm MNS} = (U^e)^{\dagger} U^\nu.
\eeq
The MNS
mixing matrix can be parametrized as
\bea
\label{MNSp}
U^{\rm MNS} &=& \pmatrix{ 1 & 0 & 0 \cr 0 & c_{23} & s_{23} \cr 0 & -s_{23} & c_{23} }
    \pmatrix{ c_{13} & 0 & s_{13}e^{-i\delta} \cr 0 & 1 & 0 \cr 
             -s_{13} e^{i\delta} & 0 & c_{13} }
    \pmatrix{ c_{12} & s_{12} & 0 \cr -s_{12} & c_{12} & 0 \cr 0 & 0 & 1 }
\nonumber \\
&=&
\pmatrix{ c_{13}c_{12} & s_{12}c_{13} & s_{13} e^{-i\delta}
\cr -s_{12}c_{23}-s_{23}s_{13}c_{12} e^{i\delta} & 
    c_{23}c_{12}-s_{23}s_{13}s_{12} e^{i\delta}
& s_{23}c_{13} \cr s_{23}s_{12}-s_{13}c_{23}c_{12} e^{i\delta}
& -s_{23}c_{12}-s_{13}s_{12}c_{23} & c_{23}c_{13}}
\eea
where $c_{ij} = \cos \theta_{ij}$
and  $s_{ij} = \sin \theta_{ij}$.
Within this parameterization, the mass-square differences for atmospheric
and solar neutrino oscillation can be chosen to be
\bea
 \Delta m^2_{\rm atm}&=&
|\Delta m^2_{32}|=|m^2_3-m^2_2|,
\nonumber \\
 \Delta m^2_{\rm sol}&=&
|\Delta m^2_{21}|= |m^2_2-m^2_1|.
\nonumber
\eea
Then the corresponding mixing angles are given by
\beq
\theta_{\rm atm}=\theta_{23},\quad
\theta_{\rm sol}=\theta_{12}, \quad
\theta_{\rm rea}=\theta_{13},
\nonumber
\eeq
where $\theta_{\rm rea}$ describes the neutrino
oscillation $\nu_\mu\rightarrow \nu_e$ in reactor
experiments such as the CHOOZ experiment.

The atmospheric neutrino data strongly suggests
near-maximal $\nu_\mu\rightarrow\nu_\tau$ oscillation with
\beq
\label{atm}
\Delta m_{\rm 32}^2 \sim 3\times 10^{-3}{\rm eV^2},
\quad
\sin^2 2\theta_{23} \sim 1.
\eeq
As for the solar neutrino anomaly, the following four
solutions are possible:
\bea
\label{sol}
&& {\rm SMA} \, : \, \Delta m_{\rm 21}^2 \sim 5.0\times 10^{-6}
{\rm eV^2},
\quad   \sin^2 2\theta_{\rm 12} \sim 2.4\times 10^{-3},
\nonumber \\
&& {\rm LMA} \, : \, \Delta m_{\rm 21}^2 \sim 3.2\times 10^{-5}
{\rm eV^2},
\quad    \sin^2 2\theta_{\rm 12} \sim 0.75,
\nonumber \\
&& {\rm LOW} \, : \, \Delta m_{\rm 21}^2 \sim 1.0\times 10^{-7}
{\rm eV^2},
\quad    \sin^2 2\theta_{\rm 12} \sim 0.96,
\nonumber \\
&& {\rm VAC} \, : \, \Delta m_{\rm 21}^2 \sim 8.6 \times 10^{-10}
{\rm eV^2},
\quad     \sin^2 2\theta_{\rm 12} \sim 0.96.
\eea
These values represent  the {\it best fit} points for each region and 
the LMA region extends to larger $\Delta m^2_{\rm 21} \sim 2\times 10^{-4}$
\cite{3gen}.
Recent reports by Super-Kamiokande \cite{SK1777} and SNO
\cite{p-sno} favor the solutions with large $\theta_{12}$.
On the other hand, the third mixing angle $\theta_{13}$ is constrained by
the CHOOZ reactor experiment \cite{CHOOZ} as
\beq
\label{choozbound}
U^{\rm MNS}_{e3} =\sin \theta_{\rm 13} \lesssim 0.2.
\eeq

The above neutrino oscillation parameters indicate that 
the neutrino mass matrix has a  nontrivial flavor structure
as the quark and charged lepton mass matrices do have.
(It has been noted that the near-maximal
atmospheric neutrino oscillation and the LMA solar neutrino
oscillation can be achieved from an anarchical neutrino mass matrix 
if one accepts certain degree of accidental cancellation
\cite{anarchy}.)
One of the most popular scheme to explain the hierarchical
quark masses and  mixing angles is the Frogatt-Nielsen mechanism 
with a spontaneously broken Abelian flavor symmetry 
\cite{fn,u1,otheru1,chunlukas,rpneut}.
In this scheme, flavor symmetry is assumed
to be broken by $\langle \phi\rangle/M_*\simeq \lambda$
($\equiv$ Cabibbo angle $\simeq 0.2$)
where $\phi$ is a symmetry-breaking scalar field and
$M_*$ denotes the fundamental scale of the model,
{\it e.g.} the Planck scale or the string scale.
Then all Yukawa couplings are suppressed by 
an appropriate power of $\lambda$ as determined by the
flavor charge of the corresponding operator, thereby
leading to  hierarchical fermion masses and mixing angles.
It is then quite natural to expect that
the nontrivial flavor structure of neutrino mass matrix
can be understood also by the Abelian flavor symmetry
explaining the hierarchical quark and charged lepton masses.

In cases of large solar neutrino
mixing, {\it i.e.} in the LMA, LOW and VAC solutions,
we have two {\it near-maximal} mixing angles $\theta_{12}, \theta_{23}$
and one {\it small} mixing angle $\theta_{13}$, as well as
the {\it small} mass-square ratio $\Delta m^2_{21}/\Delta m^2_{32}$.
It may turn out in the future neutrino experiments
that $\theta_{13}$ is significantly
smaller than the current bound (\ref{choozbound}), and then
the hierarchy between $\theta_{23}$ and $\theta_{13}$ will
become more severe.
In this paper, we wish to examine
the possibility that small $\theta_{13}$ and 
$\Delta m^2_{21}/\Delta m^2_{32}$ naturally arise together
with near bi-maximal $\theta_{23}$ and $\theta_{12}$  
as a consequence of Abelian flavor symmetry.
Our basic assumption is that the flavor symmetry is 
broken by order parameters which have the Cabibbo angle size $\lambda$.
Since the simplest scheme with single anomalous $U(1)$ flavor symmetry
and  single symmetry breaking parameter can not
produce the desired form of $M^e$ and $M^\nu$,
we need to extend the scheme.
In this regard, we consider two simple extensions, Scenario A and B,
which are assumed to be realized in supersymmetric models.
Flavor symmetry of Scenario A is a non-anomalous $U(1)_X$, 
so is broken by two scalar fields with opposite $U(1)_X$ charges $x=\pm 1$. 
In Scenario B,  flavor symmetry is extended to
$U(1)_X\times U(1)_{X^{\prime}}$ where $U(1)_X$ is anomalous while
$U(1)_{X^{\prime}}$ is non-anomalous.
It is then assumed to be broken by two scalar fields
with the flavor charges $(x,x^{\prime})=(-1,-1)$ and $(0,1)$
for which the symmetry breaking parameters naturally have the 
Cabibbo angle size.

Depending upon the way that it is generated,
$M^\nu$ can be determined either by the 
weak scale selection rule involving only the flavor charges of the
weak scale fields, or
by a more involved selection rule.
For instance, in see-saw models with
heavy singlet neutrinos $N_i$ \cite{see-saw},
the selection rule for $M^\nu$ involves
the flavor charges of $N_i$ as well as those of the weak scale
fields. 
Sometimes this feature enables us to build more
variety of models, although
in most cases it is possible to find the flavor
charges of $N_i$ for which $M^\nu$ is determined
simply by the weak scale selection rule.

Measuring the mixing angle $\theta_{13}$ is one of the main targets of
the proposed neutrino factory which can achieve
the precision down to $\theta_{13}\sim 10^{-2}$
\cite{nufac}. 
This would allow us to distinguish several different
$\theta_{13}\simeq \lambda^n$ by future
experiments. A nonzero  $\theta_{13}\simeq \lambda$ or $\lambda^2$
would give a detectable $\nu_e\leftrightarrow\nu_\mu$
transition.
On the other hand, $\theta_{13}\simeq \lambda^3$ may or may not
be detectable, and $\theta_{13}\lesssim \lambda^4$ would give
undetectably small $\nu_e\leftrightarrow\nu_\mu$ transition.
In this sense, it is meaningful to explore the possibility that
$\theta_{13}$ is as small as $\lambda^3$ or even less.
CP violating effects could
also be probed if the re-phasing invariant
$$ J_{CP}={1\over4} c_{13}^2 s_{13} \sin2\theta_{12} 
      \sin2\theta_{23}\sin\delta $$
is sizable and the 
LMA solution of the solar neutrino problem is realized \cite{nufac}.
Note that the CP violating phase is {\it not} controlled
by Abelian flavor symmetry, so
$\sin\delta$ is generically of order one in our scheme.
Another important result expected in the future neutrino experiments is 
the determination of the sign of $\Delta m^2_{32}$. 
Once $\nu_e\rightarrow\nu_\mu$ oscillations are established, matter effects 
can be measured to discriminate the sign of $\Delta m^2_{32}$ \cite{nufac}.
That is, one would be able to determine whether neutrino masses follow
the normal ($\Delta m^2_{32} >0$) or inverted ($\Delta m^2_{32} <0$)
mass hierarchy. 
As we will see, the information on $\theta_{13}$ and/or 
$\Delta m^2_{32}$ together with the solar neutrino solution 
will provide meaningful constraints on models of Abelian flavor
symmetry.

The organization of this paper is as follows.
In the next section, we discuss some aspects of 
Abelian flavor symmetry and the associated selection rule
which are relevant to our subsequent discussions.
In section III, we discuss the textures of $M^e$ and $M^\nu$
which would give small
$\theta_{13}$ and $\Delta m^2_{21}/\Delta m^2_{32}$ while
keeping the $\theta_{23}$ and $\theta_{12}$ near bi-maximal.
We focus on three type of $M^\nu$, Class (I) with 
$M^\nu_{33}\gg
M^\nu_{11}\simeq M^\nu_{12}\simeq M^\nu_{22}$ so
$m_1\simeq m_2\ll m_3$, pseudo-Dirac type Class (II) with
$M^\nu_{33}\gtrsim M^\nu_{12}\gg M^\nu_{11}, M^\nu_{22}$ so
the normal mass hierarchy $m_1\simeq m_2 \lesssim m_3$,
pseudo-Dirac type Class (III) with $M^\nu_{12}
\gg M^\nu_{11}, M^\nu_{22},M^\nu_{33}$ so
the inverted mass hierarchy $m_1\simeq m_2\gg m_3$.
In section IV, we discuss examples of
Abelian flavor symmetry for Scenarios A and B, leading to 
the mass textures discussed in section III
under the assumption that $M^\nu$ is determined by
the weak scale selection rule. 
We first list examples with largest possible $\theta_{13}$
for each of the three types of
mass textures, i.e.  Classes (I)--(III),
and the three types of solar neutrino oscillations
with large $\theta_{12}$, i.e. LMA, LOW, VAC.
We then explore the possibility
to have a smaller $\theta_{13}$.
Under the condition that the lepton doublets $L_i$ have
integer-valued flavor charges $|l_i|< 10$ when the flavor
charges of symmetry breaking fields are normalized  to
be $\pm 1$, we find the possible range of $\theta_{13}$ 
for each type of mass textures and solar neutrino oscillations
and the results are summarized in Table I.  
In section V,  we discuss see-saw models
containing  singlet neutrinos $N_i$ with integer-valued flavor charges 
$|n_i|<10$ and also with $|l_i|<10$ to find the possible range
of $\theta_{13}$. Some see-saw models are explicitly presented
as examples producing $M^\nu$ which can not be obtained under the
weak selection rule.
The results on the range of $\theta_{13}$ in see-saw models
are summarized also in Table I.
Section VI is devoted to the conclusion.


\section{Frogatt-Nielsen mechanism for Abelian flavor symmetry}

The simplest framework to implement 
the Frogatt-Nielsen mechanism with Abelian flavor symmetry
is to introduce single anomalous $U(1)_X$ 
symmetry which is assumed to be
broken by single symmetry breaking scalar field
$\langle \phi\rangle/M_*\simeq \lambda$.
This framework is best motivated from compactified heterotic string
theory with anomalous $U(1)$. In such theory, the scalar potential
includes the contribution from the string-loop induced Fayet-Illiopoulos
$D$-term, so
$$
V=\frac{g_X^2}{2}\left(\xi^2-|\phi|^2\right)^2,
$$
where $\xi^2={\rm Tr}(X)M_*^2/96\pi^2$ for the string scale
$M_*$ and all other $U(1)_X$-charged scalar fields are set
to zero for simplicity.
This framework is particularly attractive since
the symmetry breaking parameter naturally has  the Cabibbo angle size:
$$
\frac{\langle \phi\rangle}{M_*} =\left(\frac{{\rm Tr}(X)}{96\pi^2}\right)^{1/2}
\simeq \lambda.
$$
Then generic $U(1)_X$-invariant superpotential is given by
 \beq
  W = \sum_i \left({\phi\over M_*}\right)^{x_i} {\cal O}_i
\, = \, \sum_i \lambda^{x_i} {\cal O}_i \quad \quad
(x_i\geq 0),
\eeq
where the $U(1)_X$ charges of $\phi$ and ${\cal O}_i$ are
$-1$ and $x_i$, respectively.
With this selection rule, 
we can control the size of  Yukawa couplings
by assigning $U(1)_X$ charge appropriately to the low energy fields.
One important consequence of this selection rule is that the operator
${\cal O}_i$ with negative $U(1)_X$ charge is forbidden due to the
holomorphicity. 
This point is very useful and enables us to build non-trivial
Yukawa matrix.

It is well known  that realistic quark and charged lepton mass matrices
can be easily obtained within the framework
of single anomalous $U(1)_X$ and single symmetry breaking parameter
\cite{chunlukas}.
However, this framework can not provide the textures of
$M^e$ and $M^\nu$ which will be discussed in the next section
as producing bi-maximal $\theta_{23}, \theta_{12}$ 
together with small $\theta_{13},\Delta m^2_{21}/\Delta m^2_{32}$.
One simple modification of the model which would provide
the desired forms of $M^e$ and $M^\nu$
is to assume that $U(1)_X$ is 
{\it non-anomalous}, so is broken by two symmetry breaking scalar fields
$\phi_1,\phi_2$ with opposite $U(1)_X$ charges
$\pm 1$.
The $D$-term scalar potential is then given by
$$
V=\frac{g_X^2}{2}\left(|\phi_1|^2-|\phi_2|^2\right)^2
$$
which ensures 
$$
\langle\phi_1\rangle/M_*=\langle\phi_2\rangle/M_*.
$$
However there is no good reason in this framework that
$\langle\phi_1\rangle/M_*$ has the Cabibbo angle size.
A simple way to avoid this difficulty is to 
have one anomalous $U(1)_X$ and another non-anomalous
$U(1)_{X^{\prime}}$ which are broken by two scalar fields
 $\phi_1$ and $\phi_2$ having the flavor charges
$(-1,-1)$ and $(0,1)$.
In this case,  the $D$-term potential of $\phi_1$ and $\phi_2$ is given by
\beq
 V = {g_{X}^2\over 2}\left(\xi^2
-|\phi_1|^2\right)^2
 + {g_{X^{\prime}}^2\over 2}(|\phi_2|^2-|\phi_1|^2)^2\,,
\eeq
which guarantees that
\beq
\frac{\langle\phi_1\rangle}{M_*}=\frac{\langle\phi_2\rangle}{M_*}
=\frac{\xi}{M_*}\simeq \lambda.
\eeq

In this paper, we  will explore  the 
possibility of obtaining the desired textures of $M^e$ and $M^\nu$
within the following two scenarios of Abelian flavor symmetry:

\medskip
\noindent
$\bullet$ Scenario A: Single non-anomalous $U(1)_X$ 
with two symmetry breaking
parameters $\langle \phi_1\rangle/M_*
=\langle \phi_2\rangle/M_*\simeq \lambda$ with
$U(1)_X$ charges $x=\pm 1$. 
The selection rule in this scenario is given by
\beq  
W = \sum_i \lambda^{|x_i|} {\cal O}_i, \label{srvec} 
\eeq
where $x_i$ denotes the $U(1)_X$ charge of ${\cal O}_i$.

\medskip
\noindent
$\bullet$ Scenario B: \, $U(1)_{X} \times U(1)_{X^{\prime}}$ with two symmetry
breaking parameters 
$\langle \phi_1\rangle/M_*=\langle \phi_2\rangle/M_*
\simeq\lambda$ with flavor charges $(x,x^{\prime})=(-1,-1)$ and
$(0,1)$.
 The resulting selection rule is given by
\beq
W = \sum_i
\left(\frac{\phi_2}{M_*}
\right)^{x_i-x^{\prime}_i}
\left(\frac{\phi_1}{M_*}\right)^{x_i} 
{\cal O}_i
= \sum_i c_i {\cal O}_i,
\eeq
where
\beq \label{srano}
 c_i = \left\{
\begin{array}{l}
\quad 0 \quad\quad\quad\,\, \mbox{if } \, x_i<0 \,
\mbox{ or }\, x_i<x^{\prime}_i \\
 \lambda^{2x_i-x^{\prime}_i}\quad\quad \mbox{otherwise},
 \end{array} \right.
\eeq
for $(x_i,x^{\prime}_i)$ denoting the $U(1)_{X}\times
U(1)_{X^{\prime}}$ charge of ${\cal O}_i$.

The above selection rules are derived at energy scales 
just below the flavor symmetry breaking scale $M_X$.
If some heavy fields have masses
depending upon the symmetry breaking order parameter,
the low energy effective couplings of light fields 
induced by the exchange of
such heavy fields may not obey the selection rule 
as determined by the flavor charges of light fields alone.
This can happen for instance in singlet see-saw models containing 
heavy singlet neutrinos  with flavor-dependent
masses.

Usually, the smallness of neutrino masses are explained by
assuming that neutrino masses are induced by
the exchange of superheavy particles.
At the weak scale, neutrino masses are described by
$d=5$ operators in the effective superpotential:
\beq
\label{weaksuper}
\Delta W_{\rm eff}=\frac{M^\nu_{ij}}{\langle H_2\rangle^2}L_iH_2L_jH_2
\eeq
where $L_i$ ($i=1,2,3$) and $H_2$ denote the lepton and Higgs superfields,
respectively.
In singlet see-saw models,
exchanged heavy particles are the singlet
neutrinos $N_i$ having the superpotential couplings
\beq
\label{seesawsuper}
\Delta W=\frac{M^D_{ij}}{\langle H_2\rangle}
H_2L_iN_j+M^M_{ij}N_iN_j+{\rm h.c.}
\eeq
which lead to the well-known see-saw formula
 \beq \label{see-saw}
  M^{\nu}= {M^D (M^M)^{-1} (M^D)^T }.
 \eeq
Although $M^M$ and $M^D$ 
obey the selection rule as determined
by the flavor charges of the corresponding operators,
the resulting $M^\nu$ may {\it not} obey  the selection rule 
as determined by the flavor charges of
the effective operator $L_iH_2L_jH_2$.
In most cases, there exist some sets of the flavor charges
of $N_i$ for which $M^\nu$ can be determined simply by
applying the selection rule to the weak scale
effective operator $L_iH_2L_jH_2$, which
we call the weak scale selection rule (WSSR).
However it is also possible that $M^\nu$ does not
obey the WSSR, so can be determined 
only through the see-saw formula (\ref{see-saw}).

This complication does not occur in triplet see-saw models
in which $M^\nu$ is generated by the exchange of
superheavy $SU(2)_L$ triplet Higgs fields $T_1,T_2$ \cite{triplet}.
Such models include the superpotential
couplings
$$
\Delta W=h_{ij}T_1L_iL_j+h_0T_2H_2H_2+M_TT_1T_2,
$$
which give
 \beq \label{numass}
M^\nu_{ij}= h_0h_{ij} {\langle H_2 \rangle^2 \over M_T}.
 \eeq
In this case, 
$M^\nu$ can be determined always by the WSSR 
which is applied to the effective superpotential
(\ref{weaksuper}) at the weak scale.

Before closing this section, we note that
physical Yukawa couplings can be affected 
by non-holomorphic flavor-mixing terms in 
the K\"ahler potential, e.g. $\Phi_i^*\Phi_j(\phi/M_*)^{k_{ij}}$
\cite{rpneut}.
However it turns out that such K\"ahler mixing terms 
give negligible corrections
in all models discussed in this paper.

\section{Textures for bi-maximal mixing with small $U_{e3}$ }

There have been many discussions in the literatures
about the possibility of bi-maximal $\theta_{23}$ and $\theta_{12}$
\cite{bi-max1}.   Most of them rely on the assumption that $M^e$ is
(approximately) diagonal so that $U^e$ is an identity matrix.
However, comparing Eq.(\ref{MNS}) and Eq.(\ref{MNSp})
gives another interesting possibility.
If $U^e$ and $U^\nu$ are given by
\beq
\label{rotation}
U^e\simeq \pmatrix{1 & 0 & 0 \cr 0 & \frac{1}{\sqrt{2}} & \frac{1}{\sqrt{2}}
\cr 0 & -\frac{1}{\sqrt{2}} & \frac{1}{\sqrt{2}}},
\quad
U^\nu\simeq \pmatrix{\frac{1}{\sqrt{2}} & \frac{1}{\sqrt{2}} & 0 \cr
-\frac{1}{\sqrt{2}} & \frac{1}{\sqrt{2}} & 0 \cr
0 & 0 & 1},
\eeq
the resulting
$U^{\rm MNS}$ {\it naturally} has a small $\theta_{13}$ 
together with bi-maximal $\theta_{23}\sim \theta_{12}\sim \pi/4$.
In this section, we categorize what textures of $M^e$ and 
$M^\nu$ can realize this idea while giving the correct (small) value
of $\Delta m^2_{21}/\Delta m^2_{32}$.
Recall that our goal is to realize these textures within the framework
of Abelian flavor symmetry in which all mass matrix elements are
expressed in powers of the Cabibbo angle $\lambda\simeq 0.2$.
Any matrix element not shown explicitly should be understood to be 
small enough  not to disturb the basic feature of the texture.

The charged lepton mass matrix  that gives
$U^e$ of Eq. (\ref{rotation})
is given by
\beq
 M^e = m_\tau \pmatrix{\hspace{0.3cm} & \hspace{0.3cm}& \lambda^n\cr &
 & 1 \cr & & 1},     \label{tex_ch}
  \eeq
where $n \geq 1$ and the first and second column should be smaller
than the third one. 
Within the framework of Abelian flavor symmetry, there is
no way to get $U^e$ of Eq. (\ref{rotation})
other than this form of $M^e$.
However, for the
neutrino mass matrix, there are several different ways 
to get $U^\nu$ of Eq. (\ref{rotation}).
Amongst them, the following texture
with pseudo-Dirac $2\times 2$ block is of
particular interest:
\beq
M^\nu= m_{\rm max} \pmatrix{\lambda^n & \lambda^l & \cr
\lambda^l &\lambda^m & \cr
& & \lambda^k}
\eeq
where $m_{\rm max}$ denotes the largest mass eigenvalue,
$l\geq 0$, $k\geq 0$ and
$n,m > l$.
For $k=0$, this $M^\nu$ gives the normal mass hierarchy 
$m_3\gtrsim m_2,m_1$,
while $k>l=0$ gives the inverted hierarchy
$m_2\simeq m_1 \gg m_3$.
The mass eigenvalues of the above pseudo-Dirac $M^\nu$
give
\beq \label{massratio}
\frac{\Delta m^2_{\rm sol}}
{\Delta m^2_{\rm atm}}\,\sim\,
\lambda^{q+l},
\eeq
where $q \equiv {\rm min}(n,m)$.
The size of this ratio can be read off from the oscillation
data of Eqs. (\ref{atm})
and (\ref{sol}), implying
\bea  \label{lqv}
&&{\rm LMA}\,: \, q+l= 2 - 4
\nonumber \\
&&{\rm LOW}\,: \, q+l=6 - 7 
\nonumber \\
&&{\rm VAC}\,: \, q+1=9 - 10
\eea

Including the case of plain large mixing,
textures of $M^\nu$ which would give $U^\nu$ of Eq. (\ref{rotation})
together with the right value $\Delta m^2_{\rm sol}/\Delta m^2_{\rm atm}$
can be categorized as follows:

\bigskip
\noindent
$\bullet$ \, Class (I) : Plain large mixing with $n\geq 1$
which gives  $m_1\simeq m_2 \ll m_3$
 \beq  M^\nu   \label{tex_nu_A}
 \simeq m_{3} \pmatrix{
    \lambda^n & \lambda^n & \cr \lambda^n & \lambda^n &
     \cr  &  & 1} \quad \mbox{or}
 \quad M^\nu \simeq m_{3} \pmatrix{ & &
     \lambda^n \cr & & \lambda^n \cr \lambda^n & \lambda^n & 1}
 \eeq

\noindent
$\bullet$ \, Class (II) : Pseudo-Dirac type with 
$n,m>l\geq 0$ which gives the normal mass hierarchy
$m_1 \simeq m_2 \lesssim m_3$
 \beq \label{tex_nu_B}
 M^\nu \simeq
 m_3 \pmatrix{
 \lambda^n & \lambda^l & \cr \lambda^l & \lambda^m &  \cr  &  & 1}
 \eeq

\noindent
$\bullet$ \, Class (III) : Pseudo-Dirac type with the inverted
mass hierarchy 
$m_1\simeq m_2 \gg m_3$
 \beq \label{tex_nu_D} M^\nu
 \simeq m_2 \pmatrix{ \lambda^n & 1 & \cr 1 & \lambda^m &  \cr  &  & \quad
}
 \eeq
In all the cases, we will scan the possible charge assignments to find
the allowed ranges of $\theta_{13}$ which may turn out to be 
within the reach of future neutrino experiments and can give
a large CP violating quantity $J_{CP}$.  Note that Class (I) and (II)
give $\Delta m^2_{32}>0$ and Class (III) gives $\Delta m^2_{32}<0$.

\section{Models obeying the weak scale selection rule}

In this section, we discuss the models 
in which the selection rule can be applied 
to the {\it weak scale} effective superpotential:
$$
W_{\rm eff}=\frac{M^e_{ij}}{\langle H_1\rangle}H_1L_iE^c_j
+\frac{M^\nu_{ij}}{\langle H_2\rangle^2}L_iH_2L_jH_2,
$$
where $L_i,E^c_i$ and $H_1,H_2$ denote the lepton doublets, anti-lepton
singlets, and the two Higgs doublets, respectively.
As was noted in section II, this weak scale selection rule may 
not be valid in some singlet see-saw models, which will be discussed
in the next section. 
Here we consider only the models with
integer-valued flavor charges when the flavor charges
of the symmetry breaking fields are normalized to be $\pm 1$.
We further limit ourselves to the cases that
$L_i$ have the flavor charges $|l_i|<10$.
On the other hand, $E^c_i$ are 
allowed to have larger flavor charges, 
otherwise most of the LOW and VAC models presented
in the below can not be obtained.

\bigskip
\noindent
$\bullet$  Scenario A:\,
Let us first show that the neutrino mass matrix 
of  Class (I) {\it  can  not} be obtained 
under the weak scale selection rule in Scenario A.
To proceed, let  $l_i,e_i,h_1,h_2$ denote
the $U(1)_X$ charges of the superfields  $L_i,E^c_i,H_1,H_2$.
Then the charged lepton mass matrix (\ref{tex_ch}) requires
$$ 
|l_1 + e_3 + h_1| \neq |l_2 + e_3 + h_1| = |l_3 + e_3 + h_1|
$$
while the neutrino mass matrix (\ref{tex_nu_A}) requires 
$$
|l_1 + a| = |l_2 + a | \neq |l_3 + a|,
$$ 
where $a$ is a certain combination of $U(1)_X$ charges. 
These conditions inevitably lead to $M^\nu$
which {\it can not} give either a
correct value of $\Delta m^2_{\rm sol}/\Delta m^2_{\rm atm}$
or a small $\theta_{13}$.  
It appears also difficult to
find a desirable class (I) model even in the framework of singlet see-saw
models.

On the other hand, it is rather easy to get a pseudo-Dirac
$M^\nu$ of Class (II) under the weak scale selection rule.
Let us first list examples with largest possible $\theta_{13}$ for each
of the LMA, LOW and VAC solutions.
Considering the charge assignments,
 \bea  \label{model_A_2}
\mbox{LMA:} \quad\quad &l_i = (1,-2,0),\quad  e_i=(5,5,1),\quad
&h_1=h_2=0, \nonumber\\
 \mbox{LOW:} \quad\quad &l_i = (4,-7,-1),\quad  e_i=(-12,12,4),\quad
&h_1=h_2=0, \\
 \mbox{VAC:} \quad\quad &l_i = (8,-5,1),\quad  e_i=(-16,10,2),\quad
&h_1=h_2=0, 
 \nonumber
 \eea
we get the following mass textures,
 \bea  \label{matrix_A_2}
 \mbox{LMA:} \quad\quad
 \frac{M^{\nu}}{m_3}\simeq &
\pmatrix{\lambda^2 & \lambda & \lambda \cr
             \lambda & \lambda^4 & \lambda^2 \cr
             \lambda & \lambda^2 & 1 }, \quad\quad
& \frac{M^e}{m_{\tau}}\simeq 
\pmatrix{\lambda^5 & \lambda^5 & \lambda \cr
             \lambda^{2} & \lambda^2 & 1 \cr 
             \lambda^4 & \lambda^4 & 1 }\,\nonumber \\
 \mbox{LOW:} \quad\quad
 \frac{M^{\nu}}{m_3}\simeq &
\pmatrix{\lambda^6 & \lambda & \lambda \cr
             \lambda & \lambda^{12} & \lambda^6 \cr
             \lambda & \lambda^6 & 1 }, \quad\quad
& \frac{M^e}{m_{\tau}}\simeq 
\pmatrix{\lambda^5 & \lambda^{13} & \lambda^5 \cr
             \lambda^{16} & \lambda^2 & 1 \cr 
             \lambda^{10} & \lambda^8 & 1 }\, \\
 \mbox{VAC:} \quad\quad
 \frac{M^{\nu}}{m_3}\simeq &
\pmatrix{\lambda^{14} & \lambda & \lambda^7 \cr
             \lambda & \lambda^{8} & \lambda^2 \cr
             \lambda^7 & \lambda^2 & 1 }, \quad\quad
&\frac{M^e}{m_{\tau}}\simeq 
\pmatrix{\lambda^5 & \lambda^{15} & \lambda^7 \cr
             \lambda^{18} & \lambda^2 & 1 \cr 
             \lambda^{12} & \lambda^8 & 1 }, \nonumber
 \eea
for which
$$ 
(\theta_{13}, \Delta m^2_{\rm sol}/\Delta m^2_{\rm atm}) 
\simeq
(\lambda, \lambda^3)_{\rm LMA},\,\,
(\lambda, \lambda^7)_{\rm LOW}, \,\,
(\lambda^2,\lambda^9)_{\rm VAC} \,.
$$

For Class (III),  the following charge assignments are possible
 \bea  \label{model_A_3}
 \mbox{LMA:} \quad\quad &l_i = (2,-3,1),\quad  e_i=(-9,7,1),\quad
&h_1=h_2=0,\nonumber \\
 \mbox{LOW:} \quad\quad &l_i = (5,-4,2),\quad  e_i=(-13,9,1),\quad
&h_1=h_2=0, \\
 \mbox{VAC:} \quad\quad &l_i = (5,-5,-3),\quad  e_i=(-11,8,4),\quad
&h_1=h_2=0, 
 \nonumber
 \eea
to produce the mass textures
 \bea  \label{matrix_A_3}
 \mbox{LMA:} \quad\quad
 \frac{M^{\nu}}{m_2}\simeq &
\pmatrix{\lambda^3 &     1     & \lambda^2 \cr
                 1     & \lambda^5 & \lambda \cr
             \lambda^2 & \lambda & \lambda }, \quad\quad
&\frac{M^e}{m_{\tau}}\simeq 
\pmatrix{\lambda^5 & \lambda^7 & \lambda \cr
             \lambda^{10} & \lambda^2 & 1 \cr
             \lambda^6 & \lambda^6 & 1 }\,\nonumber \\
 \mbox{LOW:} \quad\quad
 \frac{M^{\nu}}{m_2}\simeq &
\pmatrix{\lambda^9 &     1     & \lambda^6 \cr
                 1     & \lambda^7 & \lambda \cr
             \lambda^6 & \lambda & \lambda^3 }, \quad\quad
&\frac{M^e}{m_{\tau}}\simeq 
\pmatrix{\lambda^5 & \lambda^{11} & \lambda^3 \cr
             \lambda^{14} & \lambda^2 & 1 \cr
             \lambda^8 & \lambda^8 & 1 }\, \\
 \mbox{VAC:} \quad\quad
 \frac{M^{\nu}}{m_2}\simeq &
\pmatrix{\lambda^{10} &     1     & \lambda^2 \cr
                 1     & \lambda^{10} & \lambda^8 \cr
             \lambda^2 & \lambda^8    & \lambda^6 }, \quad\quad
&\frac{M^e}{m_{\tau}}\simeq 
\pmatrix{\lambda^5 & \lambda^{12} & \lambda^8 \cr
             \lambda^{15} & \lambda^2 & 1 \cr
             \lambda^{13} & \lambda^4 & 1 }, \nonumber
 \eea
which give 
$$ (\theta_{13},  \Delta m^2_{\rm sol}/\Delta m^2_{\rm atm}) \simeq
(\lambda, \lambda^3)_{\rm LMA},\,\,
(\lambda, \lambda^7)_{\rm LOW},\,\, 
(\lambda^2, \lambda^{10})_{\rm VAC}. 
$$
The value of $\theta_{13}\simeq \lambda$ is
perhaps the most interesting possibility
since it is just below the current bound (\ref{choozbound}).
For the LMA and LOW,
we could easily get $\theta_{13}\simeq \lambda$ 
under the WSSR for both classes of models.
However, for the VAC solution $\theta_{13}$ can be {\it only} as large
as $\lambda^2$ under the WSSR.
As we will see in the next section,  $\theta_{13}\simeq \lambda$ 
can be obtained for the VAC in the framework
of singlet seesaw models for Class (II).

Since it may be possible to determine $\theta_{13}$ with a precision
of order $10^{-2}$,
it is worth to explore a smaller $\theta_{13}$ including
$\theta_{13}\lesssim \lambda^3$.
In this regard, the LMA in Scenario A has a special property.
Class (II) LMA models can have only $\theta_{13}\simeq \lambda$ or
$\lambda^2$, while  Class (III) LMA models can have only
$\theta_{13}\simeq \lambda$.
Actually the LMA model shown in Eq.(\ref{model_A_3}) is the unique one
which gives the LMA solution with inverted mass hierarchy.
A class (II) LMA example with $\theta_{13}\simeq\lambda^2$ is given by
\beq
\quad \quad l_i = (2,-2,0),\quad  e_i=(5,5,1),\quad h_1=h_2=0,
\nonumber 
\eeq
which lead to 
\beq
\frac{M^{\nu}}{m_3}\simeq
\pmatrix{\lambda^4 & 1 & \lambda^2 \cr
             1 & \lambda^4 & \lambda^2 \cr
             \lambda^2 & \lambda^2 & 1 }, \quad\quad
\frac{M^e}{m_{\tau}}\simeq
\pmatrix{\lambda^6 & \lambda^6 & \lambda^2 \cr
             \lambda^2 & \lambda^2 & 1 \cr
             \lambda^4 & \lambda^4 & 1 }\,. \nonumber
\eeq
Note that this form of mass matrix can give both the normal 
mass hierarchy 
($m_1 \simeq m_2 \lesssim m_3$) or the inverted mass
hierarchy ($m_1 \simeq m_2 \gtrsim m_3$)
depending on the precise values
of $M^\nu_{12}$ and $M^\nu_{33}$
both of which are of order unity.

The LOW and VAC solutions in scenario A can have 
smaller $\theta_{13}\lesssim \lambda^3$.
Here are such examples:
 \bea 
& \mbox{LOW, (II)}: \quad \quad & l_i= (3,-4,0),\quad e_i=(-10,8,2), \quad 
h_1=h_2=0,
\nonumber \\
& \mbox{VAC, (II)}: \quad\quad & l_i = (-6,4,0),\quad  e_i=(13,-8,-2),\quad
h_1=h_2=0, 
\nonumber \\
& \mbox{LOW, (III)}: \quad\quad & l_i = (4,-5,1),\quad  e_i = (-12,10,2) \quad
h_1=h_2=0, 
\nonumber \\
& \mbox{VAC, (III)}: \quad\quad & l_i = (5,-5,-1),\quad  e_i = (-12,9,3) \quad
h_1=h_2=0, 
 \eea
which produce the mass textures
 \bea  
&&\mbox{LOW, (II): } \quad\quad
 \frac{M^{\nu}}{m_3}\simeq 
\pmatrix{\lambda^6 & \lambda^1 & \lambda^3 \cr
         \lambda^1 & \lambda^8 & \lambda^4 \cr
         \lambda^3 & \lambda^4 &      1    }, \quad\quad
 {M^{e}\over m_\tau} \propto
\pmatrix{\lambda^5 & \lambda^9 & \lambda^3 \cr
         \lambda^{12} & \lambda^2 &    1    \cr
         \lambda^8 & \lambda^6 &    1    }, \nonumber \\
&& \mbox{VAC, (II): } \quad\quad
 \frac{M^{\nu}}{m_3}\simeq 
\pmatrix{\lambda^{12} & \lambda^2 & \lambda^6 \cr
             \lambda^2 & \lambda^8 & \lambda^4 \cr
             \lambda^6 & \lambda^4 & 1 }, \quad\quad
\frac{M^e}{m_{\tau}}\simeq 
\pmatrix{\lambda^5 & \lambda^{12} & \lambda^6 \cr
             \lambda^{15} & \lambda^2 & 1 \cr 
             \lambda^{11} & \lambda^6 & 1 }, \nonumber \\
&& \mbox{LOW, (III):} \quad\quad
 \frac{M^{\nu}}{m_3}\simeq 
\pmatrix{\lambda^{7} &     1        & \lambda^4 \cr
             1       & \lambda^{9}  & \lambda^3 \cr
         \lambda^{4} & \lambda^3    & \lambda }, \quad\quad
\frac{M^e}{m_{\tau}}\simeq 
\pmatrix{\lambda^5 & \lambda^{11} & \lambda^3 \cr
      \lambda^{14} & \lambda^2 & 1 \cr
      \lambda^{8} & \lambda^8 & 1 }, \nonumber  \\
&& \mbox{VAC, (III):} \quad\quad
 \frac{M^{\nu}}{m_3}\simeq 
\pmatrix{\lambda^{10} &     1        & \lambda^4 \cr
             1        & \lambda^{10} & \lambda^6 \cr
         \lambda^{4}  & \lambda^6    & \lambda^2 }, \quad\quad
\frac{M^e}{m_{\tau}}\simeq 
\pmatrix{\lambda^5 & \lambda^{12} & \lambda^6 \cr
      \lambda^{15} & \lambda^2 & 1 \cr
      \lambda^{11} & \lambda^6 & 1 }, \nonumber
 \eea
for which
$$
(\theta_{13},
\Delta m^2_{\rm sol}/\Delta m^2_{\rm atm})=
(\lambda^3,\lambda^7)_{\rm LOW,II},\,\,
(\lambda^4,\lambda^{10})_{\rm VAC,II},\,\,
(\lambda^3,\lambda^7)_{\rm LOW,III},\,\,
(\lambda^4,\lambda^{10})_{\rm VAC,III}.
$$

The examples shown in this section are the models giving either 
the largest or the smallest value of $\theta_{13}$
under the limitation $|l_i|<10$.
The reason for the occurance of these bounds on $\theta_{13}$ is that
$M^\nu_{13},M^\nu_{23}$ and $M^\nu_{11},M^\nu_{22}$  are
closely related by the $U_{X}(1)$
charge of $L_2,L_3$ fields.
It is thus difficult to suppress (enhance) $M^\nu_{13}$, $M^\nu_{23}$
arbitrarilly to get smaller (larger) $\theta_{13}$
while keeping the right size of $M^\nu_{11}$, $M^\nu_{22}$ to obtain
the right size of $m^2_{\rm sol}/m^2_{\rm atm}$ for each of 
solar neutrino oscillations.
This explains also that the VAC allows smaller $\theta_{13}$
($\lambda^2 \sim \lambda^4$) than the LMA or LOW ($\lambda \sim \lambda^3$).
The allowed ranges of $\theta_{13}$ are summarized in Table 1
for the Class (II) and (III) mass textures and the LMA, LOW, VAC
solar neutrino oscillations.

\bigskip
\noindent  
$\bullet$ Scenario B: \,
For this scenario, we use the notation that $\Phi_i(x,x^{\prime})$  
denotes the superfield $\Phi$ with $U(1)_X\times U(1)_{X^{\prime}}$
charge $(x,x^{\prime})$.
Let us first note that we need
\beq \label{mereq}
l^{\rm eff}_1 
\neq l^{\rm eff}_2 = l^{\rm eff}_3, 
\eeq
where $l_i^{\rm eff}=2l_i-l^{\prime}_i$
to get the desired form of $M^e$. 
Then, it is easy to see that Class (I)  {\it can not} 
be realized as it  requires
$l^{\rm eff}_1 = l^{\rm eff}_2 \neq l^{\rm eff}_3$.

On the other hand,  the condition (\ref{mereq}) can be reconciled
with the pseudo-Dirac structure of the Class (II) neutrino mass matrix
by imposing holomorphic zeros.  This leads us to get
the following texture:
\beq
\label{texture1}
\frac{M^\nu}{m_3}\simeq \pmatrix{\lambda^{2x} & \lambda^x & \lambda^x \cr
             \lambda^x & 0  & 0  \cr
             \lambda^x & 0  & 1 }\,,
\eeq
where $x = l^{\rm eff}_1 -l^{\rm eff}_2 = l^{\rm eff}_1 -l^{\rm eff}_3 $.
In this texture, $M^\nu_{22}$ and $M^\nu_{23}$ are forbiden due 
to the holomorphicity and the sizes of non-zero elements are 
entirely determined by the condition Eq.(\ref{mereq}).
This texture exhibits an interesting correlation of $\theta_{13}$ with 
the mass-squared difference ratio  as follows:
\beq \label{tdb2}
\theta_{13} \sim \lambda^x\,, \quad\quad
\Delta m^2_{\rm sol}/ \Delta m^2_{\rm atm} \sim \lambda^{3x}. 
\eeq
Hence $x=1,2$ or 3 is required for
the LMA, LOW or VAC, respectively,
in order to give correct square mass difference (\ref{lqv}).
We then have the following specific predictions 
$$ (\theta_{13},  \Delta m^2_{\rm sol}/\Delta m^2_{\rm atm}) \simeq
(\lambda, \lambda^3)_{\rm LMA},\,\, 
(\lambda^2, \lambda^6)_{\rm LOW},\,\,
(\lambda^3, \lambda^{9})_{\rm VAC}. $$ 
Explicit charge assignments realizing
the texture (\ref{texture1}) are given by
\bea
\label{class2}
\mbox{LMA}: \quad && L_1(0,-1), \quad L_2(1,2),\quad L_3(0,0),
\nonumber \\ && E_1(3,0),\quad E_2(2,0),\quad E_3(1,0),
\nonumber \\
\mbox{LOW}: \quad &&
L_1(0,-2),\quad L_2(1,2),\quad L_3(0,0),
\nonumber \\
&& E_1(2,-1),\quad E_2(2,0),\quad E_3(1,0),  \\ 
\mbox{VAC}: \quad &&
L_1(0,-3), \quad L_2(1,2),\quad L_3(0,0),
\nonumber \\
&& E_1(2,0), \quad E_2(2,0),\quad E_3(1,0),  \nonumber
\eea 
producing
\bea \label{matrix_B_2}
&& {\rm LMA}: \quad
\frac{M^\nu}{m_3}\simeq \pmatrix{\lambda^2 & \lambda & \lambda \cr
             \lambda & 0  & 0  \cr
             \lambda & 0  & 1 },
\quad
\frac{M^e}{m_{\tau}}\simeq \pmatrix{\lambda^5 & \lambda^3 & \lambda^1 \cr
             \lambda^4 & \lambda^2 & 1 \cr
             \lambda^4 & \lambda^2 & 1 }\, ,
\nonumber \\
&& {\rm LOW}: \quad
\frac{M^\nu}{m_3} \simeq
\pmatrix{\lambda^4 & \lambda^2 & \lambda^2 \cr \lambda^2 & 0 & 0 \cr
        \lambda^2 & 0 & 1 }, \quad
\frac{M^e}{m_{\tau}}\simeq
\pmatrix{\lambda^5 & \lambda^4 & \lambda^2 \cr 
        \lambda^3 & \lambda^2 & 1 \cr
        \lambda^3 & \lambda^2 & 1 } \, , \\
&& {\rm VAC}:\quad
\frac{M^\nu}{m_3}\simeq 
\pmatrix{\lambda^6 & \lambda^3 & \lambda^3 \cr \lambda^3 & 0 & 0 \cr
        \lambda^3 & 0 & 1 }, \quad
\frac{M^e}{m_{\tau}}\simeq \pmatrix{\lambda^5 & \lambda^5 & \lambda^3 \cr 
        \lambda^2 & \lambda^2 & 1 \cr
        \lambda^2 & \lambda^2 & 1 }\, ,  \nonumber
\eea
where $H_1$ and $H_2$ are assumed to be neutral under 
$U(1)_X\times U(1)_{X^{\prime}}$.

\medskip

Following the same argument as above, we find that 
Class (III) requires the following texture:
\beq
\frac{M^\nu}{m_3}\simeq \pmatrix{\lambda^{x} & 1 & 0 \cr
             1 & 0  & 0  \cr
             0 & 0  & 0 }\,,
\eeq
where $x = l^{\rm eff}_1 -l^{\rm eff}_2 = l^{\rm eff}_1 -l^{\rm eff}_3 $.
Here, all zero elements are again forbiden due to the holomorphicity.
This texture gives $\theta_{13}$ and the square mass difference ratio  
as
\beq \label{tdb3}
\theta_{13} \sim \lambda^x\,, \quad\quad
\Delta m^2_{\rm sol}/ \Delta m^2_{\rm atm} \sim \lambda^{x}. 
\eeq
Here we should take $x=q+l$ in Eq. (\ref{lqv})
in order to produce the right value of $\Delta m^2_{\rm sol}/
\Delta m^2_{\rm atm}$.
Then the largest possible values of $\theta_{13}$ and 
$\Delta m^2_{\rm sol}/\Delta m^2_{\rm atm}$ are
predicted to be
$$ (\theta_{13},  \Delta m^2_{\rm sol}/\Delta m^2_{\rm atm}) \simeq
(\lambda^2, \lambda^2)_{\rm LMA},\,\, 
(\lambda^6, \lambda^6)_{\rm LOW},\,\,
(\lambda^9, \lambda^{9})_{\rm VAC}. $$ 
Eq.(\ref{tdb2}) and Eq.(\ref{tdb3}) shows that the LOW and VAC solutions 
have  smaller $\theta_{13}$
than the LMA solution, and also the inverted mass hierarchy
gives  smaller $\theta_{13}$  than the normal hierarchy.
In particular, the LOW and VAC models with inverted mass hierarchy
predict  so small $\theta_{13}$ 
which can not give any observable $\nu_e\leftrightarrow\nu_\mu$
transition in the future long-base line experiments and neutrino factory.

Explicit examples of Class (III) can be  obtained by assuming
the charge assignments:
\bea
\label{class3}
{\rm LMA}: \quad &&
 L_1(0,-1),\quad L_2(0,1), \quad L_3(-1,-1),
\nonumber \\
&& E_1(3,1),\quad E_2(2,0),\quad E_3(1,0), \nonumber \\
 {\rm LOW}:\quad && 
L_1(1,-2), \quad L_2(0,2),\quad L_3(-2,-2),
\nonumber \\
&& E_1(4,5), \quad E_2(3,0),\quad E_3(2,0),  \\
{\rm VAC}:\quad && 
L_1(1,-5), \quad L_2(0,2),\quad L_3(-2,-2),
\nonumber \\
&& E_1(-1,-2),\quad E_2(3,0), \quad E_3(2,0),  \nonumber
\eea
which give
\bea  \label{matrix_B_3}
&& {\rm LMA}:\quad  \frac{M^\nu}{m_2}
\simeq \pmatrix{\lambda^2 & 1 & 0 \cr
             1 & 0  & 0  \cr
             0 & 0  & 0 }, \quad
\frac{M^e}{m_{\tau}} \simeq \pmatrix{
             \lambda^5 & \lambda^4 & \lambda^3 \cr
             \lambda^3 & \lambda^2 & 1 \cr
             \lambda^3 & \lambda^2 & 1 } 
\nonumber \\ 
&& {\rm LOW}: \quad
\frac{M^\nu}{m_{2}}\simeq \pmatrix{\lambda^6 & 1 & 0 \cr 1 & 0 & 0 \cr
        0 & 0 & 0 }, \quad
\frac{M^e}{m_{\tau}}\simeq \pmatrix{
    \lambda^5 & \lambda^{8} & \lambda^6 \cr 
            0 & \lambda^2 & 1 \cr
            0 & \lambda^2 & 1 }   \\ 
&& {\rm VAC}: \quad
\frac{M^\nu}{m_2}\simeq \pmatrix{\lambda^{9} & 1 & 0 \cr 1 & 0 & 0 \cr
        0 & 0 & 0 }, \quad
\frac{M^e}{m_{\tau}}\simeq \pmatrix{
\lambda^5 & \lambda^{11} & \lambda^{9} \cr 
        0 & \lambda^2 & 1 \cr
        0 & \lambda^2 & 1 } \nonumber
\nonumber 
\eea 
and so 
$$ (\theta_{13},  \Delta m^2_{\rm sol}/\Delta m^2_{\rm atm}) \simeq
(\lambda^2, \lambda^2)_{\rm LMA},\,\, 
(\lambda^6, \lambda^6)_{\rm LOW},\,\,
(\lambda^{9}, \lambda^{9})_{\rm VAC}. $$


 
It should be noted that 
all models discussed so far can be easily extended to the quark sector.
For instance, one can assume the following charge assignment in Scenario A 
\beq  \label{quarkA}
(q_{13},q_{23}) = (3,2),\quad 
(u_{13},u_{23}) = (5,2),\quad 
(d_{13},d_{23}) = (1,0) 
\eeq
to obtain the quark mass matrices
\beq  \label{quarkM}
\frac{M^u}{m_t}\simeq \pmatrix{ \lambda^8 & \lambda^5 & \lambda^3 \cr
         \lambda^7 & \lambda^4 & \lambda^2 \cr
        \lambda^5 & \lambda^2 & 1 } \,, \quad\quad
\frac{M^d}{m_b}\simeq   
        \pmatrix{ \lambda^4 & \lambda^3 & \lambda^3 \cr
         \lambda^3 & \lambda^2 & \lambda^2 \cr
        \lambda^1 & 1 & 1 }\,,
\eeq
where $q_{ij} = q_i - q_j, u_{ij}=u_i-u_j, d_{ij}=d_i-d_j$ 
for $q_i, u_i, d_i$ which are
the $U(1)_X$ charges of the quark superfields $Q_i, U^c_i, D^c_i$.
The same form of the quark mass matrices can be obtained
in Scenario B also from the $U(1)_X\times U(1)_{X^{\prime}}$
charge assignment:
\bea
&&Q_1(3,3), \quad Q_2(2,2), \quad Q_3(0,0),
\nonumber \\
&& U^c_1(5,5),\quad U^c_2(2,2), \quad U^c_3(0,0), \\
&&D^c_1(1,1), \quad D^c_2(0,0), \quad D^c_3(0,0). 
\nonumber 
\eea

\section{ See-saw  Models}

In singlet see-saw models, the light neutrino mass matrix is given by
 \beq \label{see-saw2}
  M^{\nu}_{ij}= \sum_{k,l} (M^M)^{-1}_{kl} M^D_{ik} M^D_{jl},
 \eeq
where $M^D$ and $M^M$ denote the Dirac and heavy-Majorana mass
matrices, respectively. 
This formula can be understood as a summation of 9 singular matrices
$M^D_{ik} M^D_{jl}$ weighted by $(M^M)^{-1}_{kl}$. 
This feature offers more variety of ways to get 
non-trivial neutrino mixing together with
hierarchical mass eigenvalues.
For example, if one contribution among the 9 contributions 
in Eq.({\ref{see-saw2})
domimates over the others, we can obtain some interesting models \cite{king}.
However, here we do not pursue this possibility, but 
look for the models without such special dominance.

\medskip
\noindent
$\bullet$ Scenario A:\,
Since the see-saw framework involves more degrees of freedom,
i.e. the flavor charges of $N_i$, one might
expect that it can reproduce all the models found under the weak
scale selection rule.
However, it is not true.
For instance, the LMA model of Class (III) in Eq.(\ref{model_A_3}) 
has no realization in see-saw framework.  Furthermore, 
it turns out that $\theta_{13}\sim \lambda$ {\it can not} be realized
in Class (III) LMA models in see-saw framework.  
On the other hand, the see-saw framework allows 
a wider range of $\theta_{13}$ than the weak scale selection rule
[see Table I] since it  provides generically a more variety
of models.
For instance, some VAC models of Class (II)
with $\theta_{13}\simeq \lambda$ can be obtained in
the see-saw framework, which was not possible
under the weak scale selection rule. 
One of such models has the flavor charges
\bea  
{\rm VAC}:\quad&&  l_i = (7,-6,-2),\quad  e_i = (-14,10,4),\quad 
 n_i = (-4,4,0)
\eea
for which the resulting $M^\nu$ and $M^e$ are given by
\bea  
{\rm VAC}:\quad && \frac{M^{\nu}}{m_3}\simeq
\pmatrix{\lambda^{10} & \lambda   & \lambda \cr
         \lambda   & \lambda^8 & \lambda^4 \cr
         \lambda   & \lambda^4 &      1     }, \quad\quad
 {M^{e}\over m_\tau} \simeq
 \pmatrix{\lambda^5 & \lambda^{15} & \lambda^9 \cr
          \lambda^{18} & \lambda^2 &    1    \cr
          \lambda^{14} & \lambda^6 &    1    }. 
 \eea
Note that one obtains completely different neutrino 
mass texture if one applies the weak scale selection rule
to the above model.


We have explored the possible range of $\theta_{13}$  under the
restriction
$|l_i|<10$ and $|n_i|<10$.
Even in  see-saw framework, it appears to be difficult
to find a desirable form of Class (I) model in Scenario A.
However there is potentially interesting example of Class (I),
yielding $\theta_{13}\simeq \lambda^2$:
\beq  \label{model_A_m}
 l_i = (2,-2,0),\quad  
 e_i = (5,5,1),\quad 
 n_i = (0,0,0)
\eeq
which  gives
\beq  
\frac{M^{\nu}}{m_3}
\simeq
\pmatrix{\lambda^4 & \lambda^4  & \lambda^2 \cr
         \lambda^4 & \lambda^4  & \lambda^2 \cr
         \lambda^2 & \lambda^2  &      1 }\, .
\eeq
The resulting
$\Delta m^2_{\rm sol}/\Delta m^2_{\rm atm}\simeq \lambda^8$
is close to either the LOW value $\lambda^{6}-\lambda^7$
or the VAC value $\lambda^9-\lambda^{10}$, so
it may fit  to the LOW or VAC if a somewhat large or small coefficient
of order one is involved.
For the LMA and LOW model of Class (II), we found that 
the range of $\theta_{13}$ is the same as the case of the
weak scale selection rule.
For the VAC of Class (II), 
$\theta_{13}\simeq \lambda$ is added
as we have noted above.
For Class (III) models, 
we find  $\theta_{13}$  can be as small as $\lambda^6$ and $\lambda^7$ for the
LMA and LOW cases, respectively.
The maximal value of $\theta_{13}$ for the LMA model of Class (III)
turns out to be of order $\lambda^2$, not of order $\lambda$,
which is noted also in the above discussion.
For the VAC model of Class (III), 
the range of $\theta_{13}$ is the same as the case of the weak scale
selection rule.
All of these results on $\theta_{13}$ are summarized in Table I.

\bigskip

\noindent
$\bullet$ Scenario B: \,
Similarly to Scenario A, 
the neutrino mass of Class (I) {\it can not} be obtained
even in the see-saw framework.
For Classes (II) and (III),
we need a pseudo-Dirac form of $M^M$ to get
a pseudo-Dirac $M^\nu$. 
We find that  all models found under the weak scale selection rule
can be realized in the see-saw framework.
For the purpose of illustration, we show only the see-saw realization 
of the LMA solution of Class (II) in Eq. (\ref{class2}).  
For this, we introduce the singlet neutrinos with 
the following $U(1)$ charges;
 \beq
    N_1(0,-1), \quad N_2(0,1), \quad N_3(0,0), \nonumber 
   \eeq
giving
 \beq
\label{seesaw2}
 M^M \propto \pmatrix{ \lambda^2 &  1 & \lambda \cr
    1 & 0 & 0 \cr \lambda & 0 & 1 }, \quad\quad
 M^D \propto 
   \pmatrix{ \lambda^2 &  1 & \lambda \cr
    \lambda & 0 & 0 \cr \lambda & 0 & 1 }.
 \eeq
The resulting $M^\nu$  is given by
 \beq
\label{seesaw2prime}
\frac{M^\nu}{m_3} \simeq \pmatrix{ \lambda^2 &
   \lambda & \lambda \cr \lambda & 0 & 0 \cr \lambda & 0 & 1 }\,,
 \eeq
which has the same form as
determined by the weak scale selection rule.

We remark that the selection rule (\ref{srano}) of Scenario B 
is very restrictive so that the see-saw framework does not provide
more freedom than the case of the weak scale selection rule.  
Basically, the positivity of the exponents
for the non-vanishing mass matrix elements 
forbids us to modify the structure of 
holomorphic zeros in the textures (37) and (41) even in the
presence of singlet neutrinos.
Therefore, no new model can be found by considering
the see-saw mechanism.

\begin{table}
\begin{center}
\caption{Possible ranges of $\theta_{13}$ for each of 
the Scenarios A and B, neutrino mass matrix of Classes (II) and (III),
and the LMA, LOW and VAC solar neutrino oscillations.
Note that Class (I) can not be obtained within our framework.
Class (II) and (III) are pseudo-Dirac type neutrino mass matrix
with $\Delta m^2_{32}>0$ and $\Delta m^2_{32}<0$, respectively.}
\begin{tabular}{|c|c|c|c|c|c|} \hline
 & solar $\nu$--oscillation & A--II & A--III & B--II & B--III \\ \hline
     &LMA  & $\lambda^2 - \lambda$ & $\lambda$ & $\lambda$ & $\lambda^2$ \\ 
 WSSR&LOW  & $\lambda^3 - \lambda$ & $\lambda^3 - \lambda$ & $\lambda^2$ & $\lambda^6$ \\ 
     &VAC  & $\lambda^4 - \lambda^2$ & $\lambda^4 - \lambda^2$ & $\lambda^3$ & $\lambda^9$ \\ 
 \hline
      &LMA  & $\lambda^2 - \lambda$ & $\lambda^6 - \lambda^2$ & $\lambda$ & $\lambda^2$ \\ 
 SEE--SAW & LOW  & $\lambda^3 - \lambda$ & $\lambda^7 - \lambda$ & $\lambda^2$ & $\lambda^6$ \\ 
      & VAC  & $\lambda^4 - \lambda$ & $\lambda^4 - \lambda^2$ & $\lambda^3$ & $\lambda^9$ \\ 
 \hline
\end{tabular}
\end{center}
\end{table}

\section{Conclusion}

In conclusion, we have examined the possibility
that the near bi-maximal mixing of atmospheric and solar
neutrinos naturally arises together with small
$U_{e3}=\sin \theta_{13}$ and $\Delta m^2_{\rm sol}/\Delta m^2_{\rm atm}$
as a consequence of Abelian  flavor symmetry.
We have considered two simple scenarios where the
mass textures are expressed in terms of 
the Cabibbo angle $\lambda$ within supersymmetric framework.
Scenario A has a  single non-anomalous $U(1)$ 
broken by two scalar fields with opposite $U(1)$ charge
and Scenario B involves one anomalous $U(1)_X$ and
another non-anomalous $U(1)_{X^{\prime}}$ which are broken
by two scalar fields with the
$U(1)_X\times U(1)_{X^{\prime}}$ charges $(-1,-1)$ and $(0,1)$.
In the latter scenario, all symmetry
breaking order parameters naturally have the Cabbibo angle size  
$\lambda \simeq 0.2$. 
Concentrating on the scheme where the large atmospheric neutrino mixing
comes from the charged lepton mass matrix,
we found that the neutrino mass 
textures of pseudo-Dirac type (with normal or inverted hierarchy) can
nicely produce a large solar neutrino mixing angle while keeping
$\theta_{13}$ appropriately small.
Current bound on $\theta_{13}$ is of order $\lambda$, however
it may be measured down to of order $\lambda^3$ in future neutrino experiments.
Table I summarizes the possible ranges of $\theta_{13}$ 
predicted by the models under consideration.  While the models of 
Scenario A produce relatively broad ranges of $\theta_{13}$,
those of Scenario B give more specific predictions which are
strongly correlated 
with $\Delta m^2_{\rm sol}/\Delta m^2_{\rm atm}$ and also
with the sign of $\Delta m^2_{32}$.  
Generically, 
larger $\Delta m^2_{\rm sol}$ come with larger $\theta_{13}$
and the normal hierarchy  ($\Delta m^2_{32}>0$) has larger
$\theta_{13}$ than the inverted hierarchy ($\Delta m^2_{32}<0$).
Table I shows that various models of neutrino mass textures
could be discriminated by future solar and terrestrial
neutrino experiments which would pin down the specific solution
of the solar neutrino problem and give information about $\theta_{13}$ and 
the sign of $\Delta m^2_{32}$.


\bigskip

{\bf Acknowledgement}:
This work is  supported by the BK21 program of Ministry of Education
(KC, EJC, WYS), Korea Research Foundation Grant
No. 2000-015-DP0080 (KC), KOSEF Grant No. 2000-1-11100-001-1 (KC),
and Center for High Energy Physics of Kyungbook National University (KC).


\end{document}